\documentclass{aa}
\usepackage{txfonts}
\usepackage{graphicx}
\usepackage{natbib}
\usepackage{amssymb}
\usepackage{lscape}
\bibpunct{(}{)}{;}{a}{}{,}

\def\chandra{{\it Chandra~}}

\def\swift{{\it Swift~}}

\def\xmm{{\it XMM-Newton~}}
\def\xmmk{{\it XMM-Newton}}

\def\m31{{M~31}}

\def\mz{HPH2011~}
\def\mzk{HPH2011}

\def\nova{{M31N~2008-05d~}}
\def\novak{{M31N~2008-05d}}

\newcommand{\nh}{\hbox{$N_{\rm H}$}~}
\newcommand{\hcm}[1]{$\times 10^{#1}$ cm$^{-2}$}

\newcommand{\tpower}[1]{$\times 10^{#1}$}
\newcommand{\power}[1]{$10^{#1}$}

\begin{document}

\title{\novak: A \m31 disk nova with a dipping supersoft X-ray light curve\thanks{Based on observations with \xmmk, an ESA Science Mission with instruments and contributions directly funded by ESA Member States and NASA.}}

\author{M.~Henze\inst{1}
	\and W.~Pietsch\inst{1}
	\and F.~Haberl\inst{1}
	\and M.~Hernanz\inst{2}
	\and G.~Sala\inst{3,4}
	\and M.~Della Valle\inst{5,6}
	\and H.~Stiele\inst{7}
}

\institute{Max-Planck-Institut f\"ur extraterrestrische Physik, Postfach 1312, Giessenbachstr., 85741 Garching, Germany\\
	email: mhenze@mpe.mpg.de
	\and Institut de Ci\`encies de l'Espai (CSIC-IEEC), Campus UAB, Fac. Ci\`encies, E-08193 Bellaterra, Spain
	\and Departament de F\'isica i Enginyeria Nuclear, EUETIB, Universitat Polit\`ecnica de Catalunya, c/ Comte d'Urgell 187, 08036 Barcelona, Spain
	\and Institut d'Estudis Espacials de Catalunya, c/Gran Capit\`a 2-4, Ed. Nexus-201, 08034, Barcelona, Spain
	\and INAF-Napoli, Osservatorio Astronomico di Capodimonte, Salita Moiariello 16, I-80131 Napoli, Italy
	\and International Centre for Relativistic Astrophysics, Piazzale della Repubblica 2, I-65122 Pescara, Italy
	\and INAF-Osservatorio Astronomico di Brera, Via E. Bianchi 46, I-23807 Merate (LC), Italy
}

\date{Received 27 April 2012 / Accepted 24 June 2012}

\abstract
{Classical novae (CNe) represent a major class of supersoft X-ray sources (SSSs) in the central region of our neighbouring galaxy \m31. Significantly different SSS properties of CNe in the \m31 bulge and disk were indicated by recent X-ray population studies, which however considered only a small number of disk novae.}
{We initiated a target of opportunity (ToO) program with \xmm to observe the SSS phases of CNe in the disk of \m31 and improve the database for further population studies.}
{We analysed two \xmm ToO observations triggered in Aug 2011 and Jan 2012, respectively, and extracted X-ray spectra and light curves.}
{We report the discovery of an X-ray counterpart to the M 31 disk nova \novak. The X-ray spectrum of the object allows us to classify it as a SSS parametrised by a blackbody temperature of $32\pm6$~eV. More than three years after the nova outburst, the X-ray light curve of the SSS exhibits irregular, broad dip features. These dips affect primarily the very soft part of the X-ray spectrum, which might indicate absorption effects.}
{Dipping SSS light curves are rarely observed in \m31 novae. As well as providing an unparalleled statistical sample, the \m31 population of novae with SSS counterparts produces frequent discoveries of unusual objects, thereby underlining the importance of regular monitoring.}


\titlerunning{\nova - \m31 disk nova with X-ray dips}

\maketitle

%
%
\section{Introduction}
\label{sec:intro}
%
Classical novae (CNe) originate from thermonuclear explosions on white dwarfs (WDs) in cataclysmic binary systems \citep[see e.g.][]{2008clno.book.....B}. Accreted hydrogen-rich material accumulates onto the WD surface under degenerate conditions until hydrogen ignition starts a thermonuclear runaway. The resulting expansion of the hot envelope increases the optical brightness of the WD dramatically within from hours to days and leads to the ejection of mass at high velocities.

During the optical nova outburst, a fraction of the accreted material can remain on the WD surface under steady hydrogen burning \citep{1974ApJS...28..247S}. This powers a supersoft X-ray source (SSS) that becomes visible, once the opacity of the ejected matter reduces sufficiently \citep{1989clno.conf...39S}. These observational \textit{SSS turn-on} timescales range from several weeks to even years \citep[][]{2007A&A...465..375P,2011ApJS..197...31S}. The duration of the SSS phase, of typically years \citep{2008ASPC..401..139K}, is limited by the amount of hydrogen fuel left on the WD and the \textit{SSS turn off} marks the cessation of the hydrogen burning \citep{2005A&A...439.1061S,2006ApJS..167...59H,2010ApJ...709..680H}.

Only X-ray observations are able to directly detect the hot WD photosphere and therefore provide important information on the physical parameters of the explosion, such as ejected and burned hydrogen mass \citep{2005A&A...439.1061S,2006ApJS..167...59H}. Estimates of the accreted and ejected masses might help us to determine the fraction of WDs in CNe that accumulate mass over time and ultimately become type Ia Supernovae.

The existence of two distinct CN populations was first suggested by \citet{1990LNP...369...34D} and \citet{1992A&A...266..232D} based on data on Galactic novae. Fast novae (which have a time of decline by two magnitudes from maximum magnitude $t_2 \leq$ 12 d) are mainly associated with the disk of the Galaxy or are concentrated close to the Galactic plane, whereas slower novae are mostly present in the bulge region of the Galaxy or at greater distances from the Galactic plane \citep{1998ApJ...506..818D}.

Novae in the central region of our neighbouring galaxy \m31 \citep[distance 780 kpc;][]{1998AJ....115.1916H,1998ApJ...503L.131S} have been found to constitute the major class of SSSs in this area \citep{2005A&A...442..879P}. Owing to its proximity to the Galaxy as well as a moderate Galactic foreground absorption \citep[\nh $\sim 6.7$ \hcm{20},][]{1992ApJS...79...77S}, \m31 is a unique target for CN surveys in the optical and X-ray regime.

Recently, \citet[][hereafter \mzk]{2011A&A...533A..52H} published a catalogue of 60 novae in \m31 with X-ray counterparts, a number that is significantly larger than for any other Galaxy, including the Milky Way. Using their catalogue, \mz were able to present the first evidence of differing X-ray properties of \m31 novae in the bulge and disk. However, small sample sizes, in particular for the relatively neglected disk novae, made those discoveries only weakly significant.

The \xmm observations this work is based on were obtained using a target of opportunity (ToO) programme aimed specifically at disk novae in \m31. Thereby, we intended to increase the knowledge about this population as well as the database for further statistical studies.

Nova \nova was discovered in the optical by \citet{2008ATel.1563....1O} on 2008-05-28.04~UT and confirmed as a nova by \citet{2008ATel.1602....1H} using H$\alpha$ observations. Furthermore, \citet{2008ATel.1567....1H} reported optical observations indicating that the object is a slow nova that still showed increasing brightness on 2008-06-09.48 UT. In this work, we report on X-ray observations of \nova with \xmmk.

%
%
\section{Observations and data analysis}
\label{sec:obs}
%
A south-east region of the \m31 disk was observed with two \xmm \citep{2001A&A...365L...1J} ToO pointings in August 2011 and January 2012 (see Table\,\ref{tab:obs_xray}). Having been scheduled at the end of their revolutions, those observations suffered from high background radiation.

We carried out the screening and processing of data from the \xmm European Photon Imaging Camera \citep[EPIC; see][]{2001A&A...365L..18S,2001A&A...365L..27T} according to the methods described in \citet{2010A&A...523A..89H} using XMMSAS \citep[\xmm Science Analysis System;][]{2004ASPC..314..759G}\footnote{http://xmm.esac.esa.int/external/xmm\_data\_analysis/} v11.0
and HEASARCs software package FTOOLS\footnote{http://heasarc.gsfc.nasa.gov/ftools/} \citep{1995ASPC...77..367B}. Spectral fitting was performed in XSPEC \citep{1996ASPC..101...17A} v12.7.0, using the T\"ubingen-Boulder ISM absorption model (\texttt{TBabs} in XSPEC), together with the photoelectric absorption cross-sections from \citet{1992ApJ...400..699B} and ISM abundances from \citet{2000ApJ...542..914W}. Additionally, data from the \xmm optical monitor \citep[OM;][]{2001A&A...365L..36M} was processed using the standard \texttt{omichain} pipeline.

%
\begin{table*}[ht]
\caption{\xmm EPIC observations of nova \nova.}
\label{tab:obs_xray}
\begin{center}
\begin{tabular}{rrrrrrrr}\hline\hline \noalign{\smallskip}
  ObsID & Exp. time$^a$ & Date$^b$ & MJD$^b$ & $\Delta t^c$ & Count Rate & L$_{0.2-1.0}$ $^d$ & $L_{\rm 50}^e$\\
  & [ks] & [UT] & [d] & [d] & [ct s$^{-1}$] & [erg s$^{-1}$] & [erg s$^{-1}$]\\ \hline \noalign{\smallskip}
  0560180101 & 20.0 (17.4) & 2008-07-18.26 & 54665.26 & 51 & $< 2.8$ \tpower{-3} & $< 0.9$ \tpower{37} & $< 1.3$ \tpower{36}\\
  0655620301 & 20.0 (6.6) & 2011-08-01.26 & 55774.26 & 1160 & ($4.0\pm0.4$) \tpower{-2} & ($1.7\pm0.2$) \tpower{38} & ($2.0\pm0.2$) \tpower{37}\\ 
  0655620401 & 30.2 (10.7) & 2012-01-21.76 & 55947.76 & 1334 & ($4.7\pm0.2$) \tpower{-2} & ($2.0\pm0.1$) \tpower{38} & ($2.2\pm0.1$) \tpower{37}\\
\hline
\end{tabular}
\end{center}
\noindent
Notes:\hspace{0.1cm} $^a $: EPIC pn total exposure time, in brackets we give the ``good`` time, corrected for dead time and high background, which was used to estimate count rates and extract the image in Fig.\,\ref{fig:xray_ima}; $^b $: Start date of the observation; $^c $: Time in days after the outburst of nova \nova in the optical on 2008-05-28.04 \citep[MJD = 54614.04; see][]{2008ATel.1563....1O}; $^d $: X-ray luminosities (unabsorbed, blackbody fit, 0.2 - 1.0 keV) and upper limits were estimated according to Sect.\,\ref{sec:results} (luminosity errors are 1$\sigma$, upper limits are 3$\sigma$); $^e $: unabsorbed equivalent luminosity computed from 0.2--10.0 keV count rates assuming a 50 eV blackbody spectrum with Galactic foreground absorption of $6.7$ \hcm{20}.\\
\end{table*}
%

\section{Results}
\label{sec:results}
\begin{figure}[t!]
	\resizebox{\hsize}{!}{\includegraphics[angle=0]{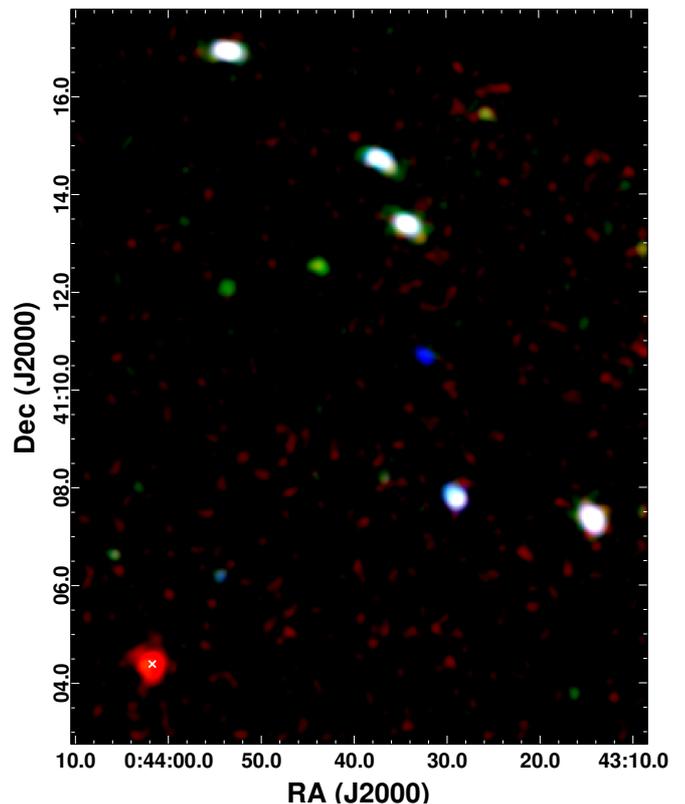}}
	\caption{\xmm EPIC RGB cut-out of observation 0655620401 showing \nova in relation to a characteristic X-ray constellation south-east of the \m31 centre. Energy bands are (0.2 -- 0.5) keV, (0.5 -- 1.0) keV, and (1.0 -- 2.0) keV for red, green, and blue channels, respectively. \nova is visible as a red source in the lower left corner of the field. The small white cross gives the position of the optical nova. The data in each colour band were binned in 2\arcsec x 2\arcsec pixels and smoothed using a Gaussian of FWHM 10\arcsec.}
	\label{fig:xray_ima}
\end{figure}

%
A new X-ray source was discovered serendipitously in the ToO observation in August 2011, which was triggered to constrain the SSS turn-off of a different \m31 disk nova (see Henze et. al. 2012, in prep.). This new source was found at RA = 00h44m01.83s, Dec = +41$\degr$04$\arcmin 23\,\farcs27$ (J2000), which is within $1\farcs3$ of the position of the CN \nova (see Fig.\,\ref{fig:xray_ima}) as given in the discovery alert by \citet{2008ATel.1563....1O}. Within their uncertainties (optical: $0\farcs3$, X-ray: $1\farcs9$), both positions are in agreement. Therefore, we identified the new source as the X-ray counterpart of the CN. No X-ray source had been known at this position from previous X-ray surveys of \m31 \citep[see][and references therein]{2011A&A...534A..55S}.

We checked archival X-ray data and found that the source was not detected during a 22~ks \xmm ToO observation on 2008-07-18.26 UT \citep[see also][]{2008ATel.1647....1P}, 51 days after the optical discovery. The corresponding upper-limit luminosity, given in Table\,\ref{tab:obs_xray}, was determined from the more sensitive EPIC pn detector. The position of the source was included in the field of view of further five X-ray observations with \chandra and \swift between this upper limit and our discovery: \chandra ACIS-I ObsIDs 11256 (20 ks; 2009-09-17.38 UT) and 11252 (59 ks; 2009-09-19.41); \swift ObsIDs 00037726002 (4.8 ks; 2008-07-22.07), 0003772503 (440 s; 2009-08-23.30), and 00037725004 (2.6 ks; 2009-09-04.56). However, the upper limits from these observations, assuming a 30~eV blackbody spectrum (see below), are at least one order of magnitude brighter than the discovery luminosity. This is caused by the short exposure times and relatively small collecting area in the case of \swift and the comparatively poor spectral response below 1~keV of the \chandra ACIS-I detector, when both are compared to \xmmk.

\begin{figure}[t!]
	\resizebox{\hsize}{!}{\includegraphics[angle=270]{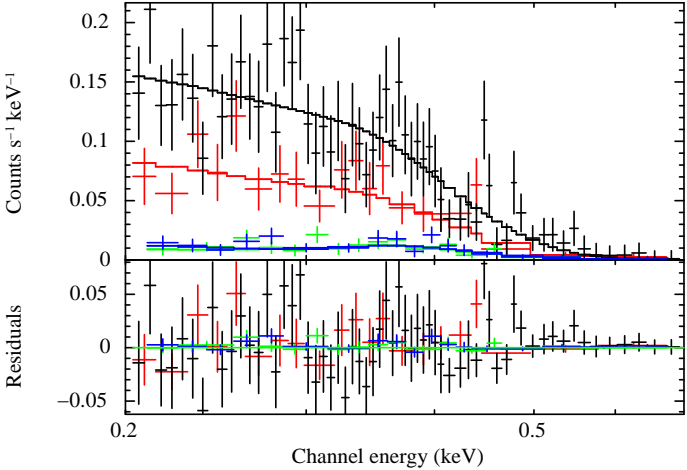}}
	\caption{\xmm EPIC spectra of \nova for obsIDs 0655620301 (pn: red) and 0655620401 (pn: black, MOS~1: green, MOS~2: blue). Blackbody fits are shown as solid lines in corresponding colours, with the lower panel giving the fit residuals.}
	\label{fig:xray_spec}
\end{figure}

The second ToO observation was triggered in January 2012, to follow up the nova counterpart which was found to still be active. The \xmm EPIC spectra of both detections are shown in Fig.\,\ref{fig:xray_spec} and compatible within their uncertainties. The merged spectrum could be described using an absorbed blackbody model with best-fit parameters $kT = 32\pm6$ eV and \nh = ($1.2^{+0.5}_{-0.4}$) \hcm{21}, allowing us to classify the source as a SSS. The SSS nature of the source provides further evidence to identify it with the optical nova. Unabsorbed luminosities inferred from the spectral model are given in Table\,\ref{tab:obs_xray}, together with ``equivalent luminosities`` ($L_{\rm 50}$) which assume a generic 50 eV blackbody spectrum with Galactic foreground absorption (\nh $\sim 6.7$ \hcm{20}) and are used, as in \mzk, to compare luminosities to earlier works \citep[e.g.][]{2007A&A...465..375P,2010A&A...523A..89H}. The blackbody parametrisation is discussed in Sect.\,\ref{sec:discuss}.

No counterpart was detected in the \xmm OM data, with upper limits of $\sim 21.5$ and $\sim 20.0$ AB magnitudes in the uvw1 and uvw2 bands, respectively, for both ToO observations.

Owing to the long gap between the only upper limit and the first detection, the SSS turn-on time of the nova is only marginally constrained to $t_{\mbox{on}} = (606\pm555)$~d. The turn-off of the SSS state is yet to happen 1334~d after the optical detection.

\begin{figure}[t!]
	\resizebox{\hsize}{!}{\includegraphics[angle=0]{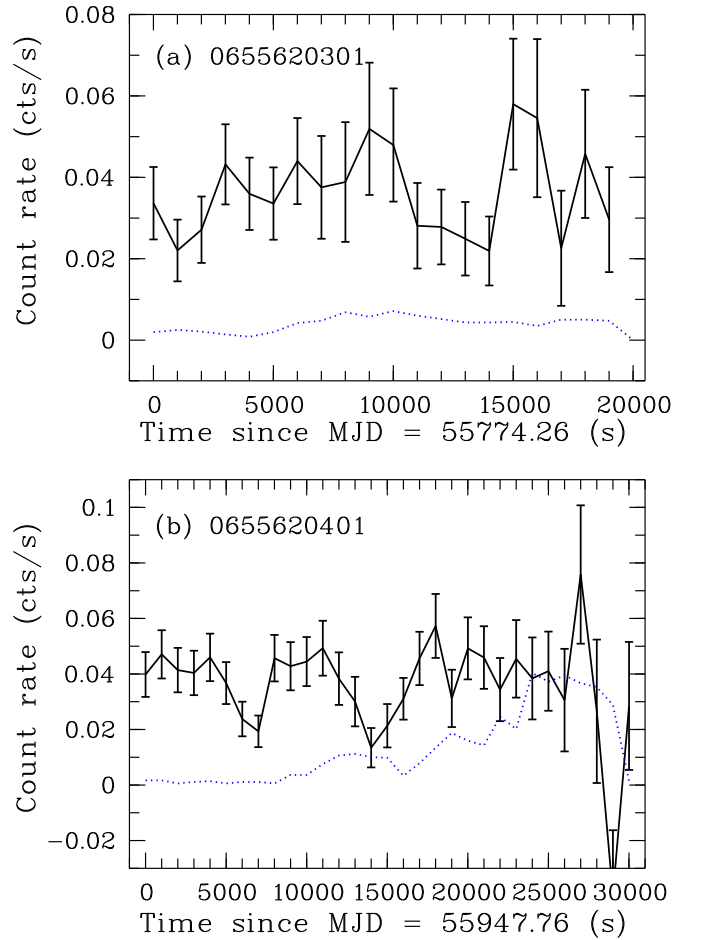}}
	\caption{Short-term, background-subtracted EPIC pn X-ray light curves of nova \nova in the 0.2--1.0~keV range during \xmm observations 0655620301 (upper panel, a) and 0655620401 (lower panel, b) with 1000~s binning and $3\sigma$ error bars. In blue, the area-corrected, background light curves in the same energy range are shown for a source-free region close to the nova.}
	\label{fig:xray_lc}
\end{figure}

The X-ray short-term light curve of nova \nova displayed variability during both ToO pointings (Fig.\,\ref{fig:xray_lc}). The first observation (ObsID 0655620301) showed a relatively stable count rate during the initial $\sim 10$~ks, after which there are indications for a dip that lasted $\sim 4$~ks. After this possible dip, during which the count rate dropped by a factor of $\sim 2$, the count rate seemed to return to the previous level. Although the ingress of the dip shows a certain resemblance to the soft-background light curve, its general shape cannot be explained by the background alone (see Fig.\,\ref{fig:xray_lc}(a)). 

In the second observation, the dip features are considerably more significant. Two pronounced dips can easily be identified in the first 20~ks of the light curve, as shown in Fig.\,\ref{fig:xray_lc}(b). While the first dip, by a factor of $\sim 2$, lasted for $\sim 3$~ks and appeared asymmetric, the second, possibly deeper dip had a more symmetric profile and was $\sim 5$~ks long.

The first dip happened in its entirety during the initial $\sim 8$~ks of the observation, which were characterised by low background (see the blue background light curve in Fig.\,\ref{fig:xray_lc}(b)). In contrast, the second dip took place during a time of higher background, but shows neither suspicious symmetry nor anti-symmetry with the background light curve. Measurements during the last $\sim 7$~ks of the observation are unreliable owing to the presence of a very high background and are only shown for completeness. 

We found that the possible dip during the first observation is only mildly significant within $(1-2)\sigma$ with respect to the rest of the light curve. In constrast, both dips during the second ToO pointing exceed a $3\sigma$ significance level in two bins each. This difference might be connected to the otherwise stable brightness level of \nova during the second observation, but it ultimately means that the evidence of a dipping SSS light curve  is solely based on this second observation, with the first light curve only providing a hint about this similar previous behaviour. The identified dip features can be attributed to the nova itself and are unlikely to stem from background variations. No periodicities were found in either light curve within the range of \power{-4} - \power{-1} Hz using the HEASoft \texttt{powspec} tool.

\section{Discussion}
\label{sec:discuss}
%
\nova was discovered in the south-eastern part of the \m31 disk. Although its position lies broadly along the minor axis of the galaxy, the nova is sufficiently far away from the bulge for projection effects to have no effect on its classification as a disk nova. The object does not appear to be associated with a bright spiral arm of \m31.

The relationships between the X-ray parameters of \nova follow the trends found in the \m31 sample of novae with SSS counterparts by \mzk. The long SSS phase together with the low blackbody temperature point towards a slowly evolving nova on a low-mass WD \citep[see][\mzk]{2010ApJ...709..680H}. This interpretation also fits the slow evolution in the optical reported by \citet{2008ATel.1567....1H}. Although \m31 disk novae are suspected to be on average hotter SSSs (mean blackbody $kT = 56$~eV in \mzk) than bulge novae, the temperature derived for \nova lies within the low-energy tail of the current, relatively broad temperature distribution (see figure 13 in \mzk).

Unfortunately, the OM non-detection does not provide additional constraints on the underlying binary. In the optical blue/ultraviolet regime, the emission of a CN system in quiescence is dominated by the accretion disk rather than by either the main-sequence or red-giant secondary star \citep[e.g.][]{2012ApJ...746...61D}. While \nova is not yet in quiescence, the emission from the WD photosphere has been effectively shifted out of the optical/ultraviolet range \citep[see][]{2006ApJS..167...59H}, thereby creating the SSS. Owing to the distance of \m31, our OM upper limits do not provide sufficient sensitivity to distinguish between the CN secondary star classes \citep[compare also][]{2009ApJ...705.1056B}.

\begin{figure}[t!]
	\resizebox{\hsize}{!}{\includegraphics[angle=270]{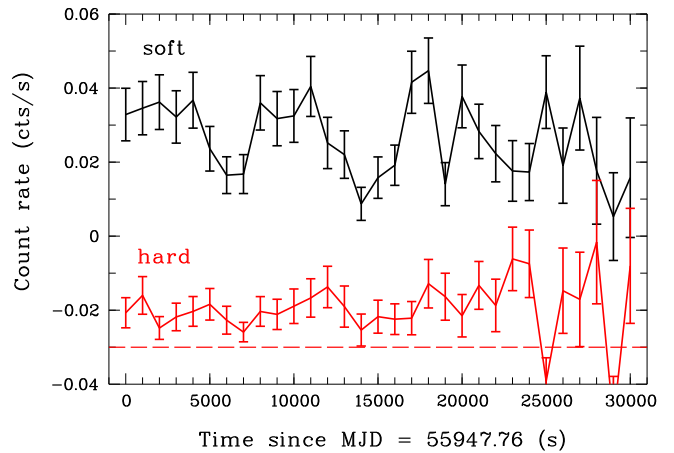}}
	\caption{Same as Fig.\,\ref{fig:xray_lc}(b) for energy bands 0.2--0.35~keV (soft, black) and 0.35--0.6~keV (hard, red). The lower light curve includes an offset zero level (dashed red line) for better readability.}
	\label{fig:xray_hard}
\end{figure}

Novae with dipping short-term X-ray light curves are rare in \m31. The only comparable object was nova M31N~2006-04a, whose SSS light curve displayed three deep minima during a 20~ks \xmm observation \citep{2010A&A...523A..89H}. However, \citet{2010A&A...523A..89H} reported for M31N~2006-04a an apparent periodicity, which appeared to indicate the binary period of the system. In contrast, the aperiodic nature and different shape of the dips observed in \nova make an association with the WD rotation or the orbital period of the binary system unlikely. To investigate the nature of the dips further, we divided the light curve of observation 0655620401 (see Fig.\,\ref{fig:xray_lc}(b)) into two energy bands based on the SSS spectrum: 0.2--0.35~keV and 0.35--0.6~keV. Above 0.6~keV, we detected no photons from the source. The resulting light curves for the relatively soft and hard ranges are shown in Fig.\,\ref{fig:xray_hard}.

The figure illustrates that the dips are almost exclusively confined to energies below 0.35~keV. This is a strong hint that the dips might be caused by photoelectric absorption in intervening material, which has a smaller effect at higher energies. In this scenario, the material, possibly gas clouds, would be crossing the line of sight randomly. The different length and shape of the individual dips suggests that the absorbers have wide ranges of sizes and/or velocities. We also extracted and fitted two separate spectra for the dipping and non-dipping times during observation 0655620401. However, although we found that the best-fit \nh is larger during the dips by about a factor of two (1.9 vs 1.0 \hcm{21}), owing to the low count rates the errors are large and both extinction values are still compatible on the $1\sigma$ level.

However, high-resolution spectra of Galactic novae \citep[e.g.][]{2008ApJ...673.1067N,2011ApJ...733...70N,2012ApJ...745...43N} emphasize that blackbody fits only provide a qualitative parametrisation of SSS spectra and not a physically realistic model \citep[see also][]{1991A&A...246L..17G,1997ARA&A..35...69K}. Therefore, the variability of \nova could also originate in the changing properties of individual emission or absorption features that are not considered by our simple model. One example for such an effect is the explanation of variability caused by a variation in the optical depth of the O\,{\sc i} absorption edge in the SSS state of the Galactic nova RS~Ophiuchi \citep{2007ApJ...665.1334N}.

While dipping SSS light curves are rare in \m31 novae, Galactic novae have been observed to show dip-like or eclipse features in X-rays on relatively short timescales. Most of these objects behave however, in a significantly different way from \nova \citep[see e.g.][]{2003ApJ...594L.127N,2008ApJ...675L..93S,2011ApJ...733...70N}. The only exception is the recurrent nova U~Sco, for which similar dips to those in Fig.\,\ref{fig:xray_lc}(b) were found in its 2010 outburst by \citet{2012ApJ...745...43N}. These authors note that the X-ray dips, shown in their figure 2, could originate from absorption by ``clumpy absorbing material that intersects the line of sight while moving along highly elliptical trajectories``. Interestingly, \citet{2012ApJ...745...43N} report the dips to be present only in an early state of the SSS phase (22.9~d after outburst) and to have already disappeared in a second observation on day 34.9, a behaviour that they interpret as a signature of ``significant progress in the formation of the [accretion] disk.''

If a similar picture applied to \novak, the timescales involved would be considerably longer, since the \m31 nova experienced its dips at a much later stage: more than three years after outburst. However, \nova shows an overall slower evolution in X-rays and optical than U~Sco. Little is known to date about how quickly CN accretion disks reform \citep[e.g.][]{1998MNRAS.293..145R,2002Sci...298..393H,2012ApJ...745...43N}. \nova might in respect of disk reformation resemble U~Sco on longer timescales. In this case, the high-inclination nature of the U~Sco binary system \citep[$83\degr \pm 3\degr$,][]{2001MNRAS.327.1323T} suggests that we might also be viewing \nova at high inclination.

An alternative scenario to explain the dips could be absorption in the ejected nova shell. Inhomogeneous and clumpy ejecta have been observed for several Galactic novae in their resolved nebular remnants \citep[e.g.][]{2000MNRAS.314..175G}, optical spectra \citep[e.g.][]{1991ApJ...376..721W}, and even in X-rays \citep[e.g.][]{1999ApJ...518L.111B}. Evidence of density inhomogeneities in the ejected material was also discussed for U~Sco based on optical data \citep{2010AJ....140.1860D}. \citet{2012ApJ...745...43N} ruled out the possibility that the dips they found in U~Sco were caused by the inhomogeneous ejecta, on the basis that these clumps are expected to be highly ionized and therefore should not obscure soft X-rays via photoelectric absorption. With regard to the later evolutionary stage of \novak, this argument might not, however hold in our case. Furthermore, the possibility of Thomson scattering within the hot ejecta was also rejected by \citet{2012ApJ...745...43N} for U~Sco, as it would have led to optical/UV dips detectable in their simultaneous \xmm OM fast-mode light curves. For \novak, owing to the OM non-detection and novae in \m31 being more distant, this cross check is unfortunately impossible. However, the energy dependence shown in Fig.\,\ref{fig:xray_hard} is strongly indicative of photoelectric absorption.

Another possible scenario might be the occultation of the WD by cold, clumpy material caused by the impingement of a high-accretion rate stream on the accretion disk, as discussed by \citet{1997A&A...318...73S} for the canonical SSS CAL~87. However, applying this model as constructed for CAL~87 to \nova would result in a permanent attenuation of the WD by a kind of curtain of spray material. While this curtain would extend beyond the accretion disk towards a large vertical height, and thereby permit a larger range of inclination angles, the permanence of the occultation differs decidedly from the distinct dips we observed for our object. Nevertheless, it is possible that an inhomogeneously extended accretion stream or stream-impact spray could cause the observed non-periodic X-ray variability. Therefore, the model of \citet{1997A&A...318...73S} might help to explain the light curve of \nova within a more sophisticated framework, the construction of which is beyond the scope of this paper.

The high nova rate and the proximity of \m31 have not only allowed us to construct an unparalleled statistical sample but also provided us with a number of unusual objects. Both features enable our understanding to advance by revealing patterns and exceptions within the nova population. A continuing X-ray monitoring of \m31 is therefore strongly needed to provide a sufficiently large database for future discoveries.

\begin{acknowledgements}
We thank the anonymous referee for her/his constructive comments and suggestions. The X-ray work is based in part on observations with \xmmk, an ESA Science Mission with instruments and contributions directly funded by ESA Member States and NASA. The \xmm project is supported by the Bundesministerium f\"{u}r Wirtschaft und Technologie / Deutsches Zentrum f\"{u}r Luft- und Raumfahrt (BMWI/DLR FKZ 50 OX 0001) and the Max-Planck Society. M. Henze acknowledges support from the BMWI/DLR, FKZ 50 OR 1010. GS acknowledges the MICINN grants AYA2010-15685 and AYA2011-23102, and the ESF EUROCORES Program EuroGENESIS through the MICINN grant EUI2009-04167. M. Hernanz acknowledges support from the MICINN grant AYA2011-24704. 

\end{acknowledgements}

\bibliographystyle{aa}

\end{document}